\documentclass[usenatbib,usegraphicx,onecolumn]{mn2e}
\usepackage{amsmath,aas_macros,multicol}

\defcitealias{sls04}{SLS04}
\defcitealias{vwo+04}{V04}
\defcitealias{whl+04}{W04}
\defcitealias{wvw+06}{W06}
\defcitealias{tm06}{TM06}

\newcommand{\citeta}[1]{\citetalias{#1}}
\newcommand{\citepa}[1]{\citepalias{#1}}

\newcommand{\integral}{\textit{INTEGRAL}}
\newcommand{\xmm}{\textit{XMM}}
\newcommand{\xmmlong}{\textit{XMM-Newton}}
\newcommand{\bepposax}{\textit{Beppo-SAX}}
\newcommand{\konuswind}{\mbox{Konus-WIND}}
\newcommand{\swift}{\textit{Swift}}

\def\arcdeg{\hbox{$^\circ$}}
\def\slantfrac#1#2{\hbox{$\,^#1\!/_#2$}}
\def\onehalf{\slantfrac{1}{2}}

\newcommand{\keV}{\mbox{$\rm\,keV$}}

\newcommand{\um}{\mbox{$\mu$m}}

\newcommand{\tee}[1]{\mbox{$\times 10^{#1}$}}

\newcommand{\xray}{\mbox{X-ray}}
\newcommand{\gray}{\mbox{gamma-ray}}

\newcommand{\grbone}{\mbox{GRB\,031203}}
\newcommand{\grbtwo}{\mbox{XRF\,060218}}

\newcommand{\sngrb}{\mbox{SN\,1998bw}}
\newcommand{\sngrbb}{\mbox{GRB\,980425}}
\newcommand{\xrfsn}{\mbox{XRF\,020903}}
\newcommand{\grbsn}{\mbox{GRB\,030329}}
\newcommand{\grbflare}{\mbox{GRB\,050502B}}
\newcommand{\snxrt}{\mbox{SN\,2008D}}

\newcommand{\ergcms}{\mbox{erg cm$^{-2}$ s$^{-1}$}}
\newcommand{\ergcmsq}{\mbox{erg cm$^{-2}$}}

\newcommand{\hyperfsym}{\mbox{$_2F_1$}}
\newcommand{\hyperf}[4]{\mbox{$_2F_1(#1,#2,#3,#4)$}}

\newcommand{\beq}{\begin{equation}}
\newcommand{\eeq}{\end{equation}}

\newcommand{\url}[1]{{\tt #1}}
\newcommand{\email}[1]{{\tt #1}}
\newcommand{\phm}[1]{\phantom{#1}}

\newcommand{\HI}{\hbox{{\rm H}\kern 0.1em{\sc i}}}

\def\simlt{\mathrel{\hbox{\rlap{\hbox{\lower4pt\hbox{$\sim$}}}\hbox{$<$}}}}
\def\simgt{\mathrel{\hbox{\rlap{\hbox{\lower4pt\hbox{$\sim$}}}\hbox{$>$}}}}

\def\ale{\mathrel{\hbox{\rlap{\hbox{\lower4pt\hbox{$\sim$}}}\hbox{$<$}}}}
\def\age{\mathrel{\hbox{\rlap{\hbox{\lower4pt\hbox{$\sim$}}}\hbox{$>$}}}}

\begin{document}

\title[GRB\,031203 and XRF\,060218 as Cosmic Twins]%
   {Could the GRB-Supernovae GRB\,031203 and XRF\,060218 be Cosmic Twins?}

\author[Feng \& Fox]%
   {Lu Feng\thanks{Current address:  Department of Physics,
                26-524, Massachusetts Institute of Technology,
                Cambridge, MA 02139, USA; \email{lufeng@mit.edu}}
    \& Derek B. Fox \\
    Department of Astronomy \& Astrophysics, 
    525 Davey Laboratory, 
    Pennsylvania State University,
    University Park, PA 16802,
    USA; \\
    \email{lwf5001@astro.psu.edu; dfox@astro.psu.edu}}

\maketitle


\begin{abstract}

The \gray\ burst (GRB) / \xray\ flash (XRF) events \grbone, discovered
by \integral, and \grbtwo, discovered by \swift, represent two of only
five GRB-SNe with optical spectroscopic confirmation of their SN
components.  Yet their observed high-energy properties offer a sharp
contrast: While \grbone\ was detected as a short $40$-s burst with a
spectrum peaking at $E_{\rm peak}>190$\keV, \grbtwo\ was a
$T_{90}\approx 2100$-s long, smoothly evolving burst with peak energy
$E_{\rm peak}=4.9$\,keV.  At the same time, the properties of the two
expanding dust-scattered \xray\ halos observed in a fast-response
\xmmlong\ observation of \grbone\ reveal that this event was
accompanied by an ``\xray\ blast'' with fluence comparable to or
greater than that of the prompt \gray\ event.  
Taking this observation as our starting point, we investigate the
likely properties of the \xray\ blast from \grbone\ via detailed
modeling of the \xmm\ data, discovering a third halo due to scattering
off a more distant dust sheet at $d_3 = 9.94\pm 0.39$\,kpc, and
constraining the timing of the \xray\ blast relative to the GRB
trigger time to be $t_0 = 11\pm 417$\,s.
Using our constraints, we compare the properties of \grbone\ to those
of other GRB-SNe in order to understand the likely nature of its
\xray\ blast, concluding that a bright \xray\ flare, as in \grbflare,
or shock breakout event, as in \grbtwo, provide the most likely
explanations.  In the latter case, we consider the added possibility
that \grbtwo\ may have manifested an episode of bright \gray\ emission
prior to the burst observed by \swift, in which case \grbone\ and
\grbtwo\ would be ``cosmic twin'' explosions with nearly identical
high-energy properties.

\end{abstract}



\begin{multicols}{2}

\section{Introduction}
\label{sec:intro}

The discovery of the connection between long-duration \gray\ bursts
(GRBs) and \xray\ flashes (XRFs), on the one hand, and Type~Ibc
supernovae (SNe) on the other, has confirmed the collapsar model as
proposed by \citet{w93} -- for a recent review, see \citet{wb06}.  

Several distinct types of investigation provide support for the GRB-
(XRF-) SN connection; however, the most convincing evidence has come
from spectroscopic observations of coincident supernovae associated
with the lowest-redshift GRBs and XRFs.  As is frequently the case in
astronomy, these nearby and best-studied cases have each proven
anomalous in their own way; even considered as a group, the properties
of GRB-SNe thus resist ready generalisation.

In all, five spectroscopically confirmed GRB-SNe -- a term we will use
to refer to the GRB- and XRF-SNe, as a group -- have been observed
to-date.  \sngrb\ offered the first direct evidence for GRB-SN
association, as the SN that appeared in the error box of GRB\,980425,
with an explosion time consistent with the GRB trigger, was shown to
be exceptionally luminous and energetic in the optical \citep{gvv+98},
with an associated mildly relativistic ($\Gamma\simgt 3$) outflow that
made it uniquely radio-bright \citep{kfw+98}.  Occurring in a
catalogued galaxy at a distance of just 36\,Mpc ($z=0.0085$), however,
GRB\,980425 was four orders of magnitude less luminous than the
typical $z\simgt 1$ burst.

\xrfsn, at $z=0.25$ the first XRF with a determined redshift and
observed optical and radio afterglow emission \citep{skb+04a}, also
provided the first XRF-associated supernova \citep{skf+05,bfs+06}; its
spectrum was ultimately shown to be similar to that of \sngrb\ 
\citep{skf+05}, confirming the deep connection between XRFs and GRBs.

\grbsn\ / SN\,2003dh at $z=0.17$ provided the first definitive
association of a ``cosmological'' GRB with a coincident type~Ic
supernova, since it had an isotropic-equivalent \gray\ output of
$E_{\gamma{\rm ,iso}}=10^{52}$\,erg \citep{vsb+04}, within the
observed range for $z\simgt 1$ events.  Broad-line SN~Ic features were
visible within a few days of the GRB \citep{smg+03}, and the ultimate
evolution of the supernova showed a striking similarity to \sngrb\ 
\citep{hsm+03,mgs+03}, validating the two earlier associations.

Later that year, \grbone\ was detected by \integral\ with a duration
of $40$-s and a peak energy of $>190$\keV\ \citep[][hereafter
  \citeta{sls04}]{sls04}.  Six hours after the burst, a fast-response
\xmmlong\ observation of the burst position began; a fading
\xray\ afterglow was quickly identified \citep{gcn.2474}, coincident
with a $z=0.105$ galaxy \citep{pbc+04}.  This was confirmed as the
host galaxy once an associated type~Ic supernova, SN\,2003lw, was
discovered \citep{mtc+04}.  Investigation of the radio and
\xray\ energetics of the event revealed further similarities to
SN\,1998bw \citep{skb+04b}, although the SN itself was somewhat
subluminous by comparison \citep{gmf+04}.

Contemporaneously, analysis of the \xmm\ data revealed two expanding
\xray\ halos centred on the fading GRB afterglow \citep[][hereafter
  \citeta{vwo+04}]{vwo+04}; these halos were convincingly interpreted
as scattered photons from a bright \xray\ ``blast'' that occurred
near-simultaneously with \grbone.  Modeling the properties of the
\xray\ blast, \citet[][or \citeta{whl+04}]{whl+04} and \citet[][or
  \citeta{wvw+06}]{wvw+06} argued that the prompt emission of
\grbone\ was, in fact, an \xray\ flash, exhibiting greater fluence in
\xray\ compared to \gray, $\log
(S_{\text{2--30\,keV}}/S_{\text{30--400\,keV}}) = 0.6\pm 0.3 > 0$.
Independent analysis by \citet[][or
  \citeta{tm06}]{tm06} yielded a lower \xray\ fluence estimate which
nonetheless also satisfies $S_{\text{2--30\,keV}} > S_{\text{30--400\,keV}}$.

The nature of \grbone\ is not easily resolved, for two reasons.
First, the bright soft \xray\ emission cannot be interpreted as the
low-energy tail of \grbone, as its fluence significantly exceeds the
low-energy extrapolation of the \integral\ spectrum
\citepa{sls04,wvw+06,tm06}.  Second, the \integral\ observations
themselves would have detected the \xray\ blast if it was emitted over
a brief time interval prior to or within 300\,s following the burst
itself, before 22:06:27 UT, when \integral\ began to slew to a new
pointing.  Thus, two distinct explosive episodes are required, with
the \xray\ blast most likely occurring both physically and temporally
apart from the GRB.  To satisfy these constraints, \citeta{wvw+06}
consider the possibility that \grbone\ was followed by a bright
\xray\ flare such as that observed by \swift\ from \grbflare\ 
\citep{brf+05}; such flares -- albeit rarely as bright as that case --
are familiar features of the GRB \xray\ lightcurves now routinely
gathered by \swift\ \citep[e.g.,][]{cmr+07}.

\grbtwo, detected by \swift\ \citep{cmb+06} and coincident with a
$z=0.033$ host galaxy \citep{mha+06}, evolved into a type~Ic supernova
that was observed from its earliest moments, including extensive
spectroscopic observations \citep{mha+06,pmm+06,mdn+06}, thanks to the
prompt alert and extended follow-up observations of \swift.  Analysis
of the \xray, optical, and radio energetics of \grbtwo\ demonstrate a
deep similarity to \sngrb\ and \grbone\ \citep{skn+06}; however,
\grbtwo\ remains an anomaly in that its prompt emission exhibited a
uniquely extended evolution, having a 90\%-containment duration of
$T_{90}=2100$\,s and a correspondingly low peak flux $F_{\text{peak}}
= 8.8\tee{-9}$\,\ergcms\ (15--150 keV).  This peak flux is more than
an order of magnitude lower than those of all other GRB-SNe except for
\xrfsn, which had an even softer spectrum; indeed, it was only
detected because of the ``image trigger'' capability of the
\swift\ BAT, which identified \grbtwo\ as an uncatalogued source
apparent in the first 64\,s integration taken at that pointing.

Emission from \grbtwo\ is present in the first image taken by
\swift\ on that particular orbit; as such, a GRB similar to
\grbone\ could have occurred prior to the start of
\swift\ observations.  Moreover, a significant fraction of the prompt
\xray\ emission of \grbtwo\ is supplied by a thermal component,
unobserved in other GRB-SNe, that is attributed to the shock breakout
of the associated supernova \citep{cmb+06} rather than to the
relativistic jet usually invoked to explain the prompt emission.  If
the \xray\ blast from \grbone \ can be explained as a similar shock
breakout, then these two events -- despite their very different
appearances -- would in fact be ``cosmic twins'' exhibiting nearly
identical evolution in the \xray\ and \gray\ bands (see also
\citealt{ggm+06}).

In this paper, we undertake a quantitative exploration of this
hypothesis by constraining the properties of the \xray\ blast in
\grbone\ via detailed modeling of its expanding \xray\ halos; this
analysis also allows us to address the alternate theory that the
\xray\ emission from \grbone\ was due to a later \xray\ flare
\citepa{wvw+06}. \S2 presents the \xmm\ observations of the afterglow
and halos of this burst and the details of our analysis, including
derivation of parameter uncertainties.  \S3 discusses our results in
light of various hypotheses regarding \grbone\ and \grbtwo, and places
these events in the larger context of the known varieties of cosmic
explosion.  \S4 summarizes our conclusions.


\section{Observation and Data Analysis}
\label{sec:obs}

The GRB\,031203 data (Observation ID 0158160201), downloaded from the
\xmmlong\ archive, are processed and reduced for EPIC MOS and PN
cameras through the standard procedures (tasks {\tt emchain} and {\tt
  epchain}), with standard filtering, using the \xmm\ Science Analysis
System (SAS), version 7.1.0.  The observation had start and stop times
of 04:10:18 UT and 20:16:20 UT on 4 Dec 2003, which after standard
filtering yields 57.8\,ksec of good time and 57.3\,ksec net exposure
in CCD~1 of EPIC-MOS1 and MOS2, and 56.2\,ksec of good time and a
CCD-dependent net exposure between 53.4 and 54.0\,ksec in EPIC-PN.
The observation was not affected by significant soft-proton flaring.  


\subsection{Model Construction and Fitting}
\label{sub:obs:model}

We consider a circular region with radius 311\arcsec\ centred on the
GRB afterglow, and restrict our attention to detected counts in the
energy range 0.2 to 3.0\,keV (0.2 to 4.0\,keV for PN).  We generate an
exposure map with 1\arcsec\ pixel resolution, using the {\tt eexpmap}
task, to account for PN active regions; exposure map values are
treated as binary (on or off) in the subsequent analysis, and no
initial exposure map is created for the MOS data as our circular
region of interest is fully contained on the central CCD.  Bright
\xray\ sources within the region of interest are identified and all
counts within an exclusion radius of 15\arcsec\ of each source
position are excluded from analysis; these exclusion regions are then
integrated into exposure maps for the PN and MOS data for analysis
purposes.  After all exclusions we accept for analysis 10,754 counts
from the MOS-1 detector, 10,535 counts from MOS-2, and 32,085 counts
from PN.

\begin{figure*}
\centerline{\includegraphics[height=50mm]{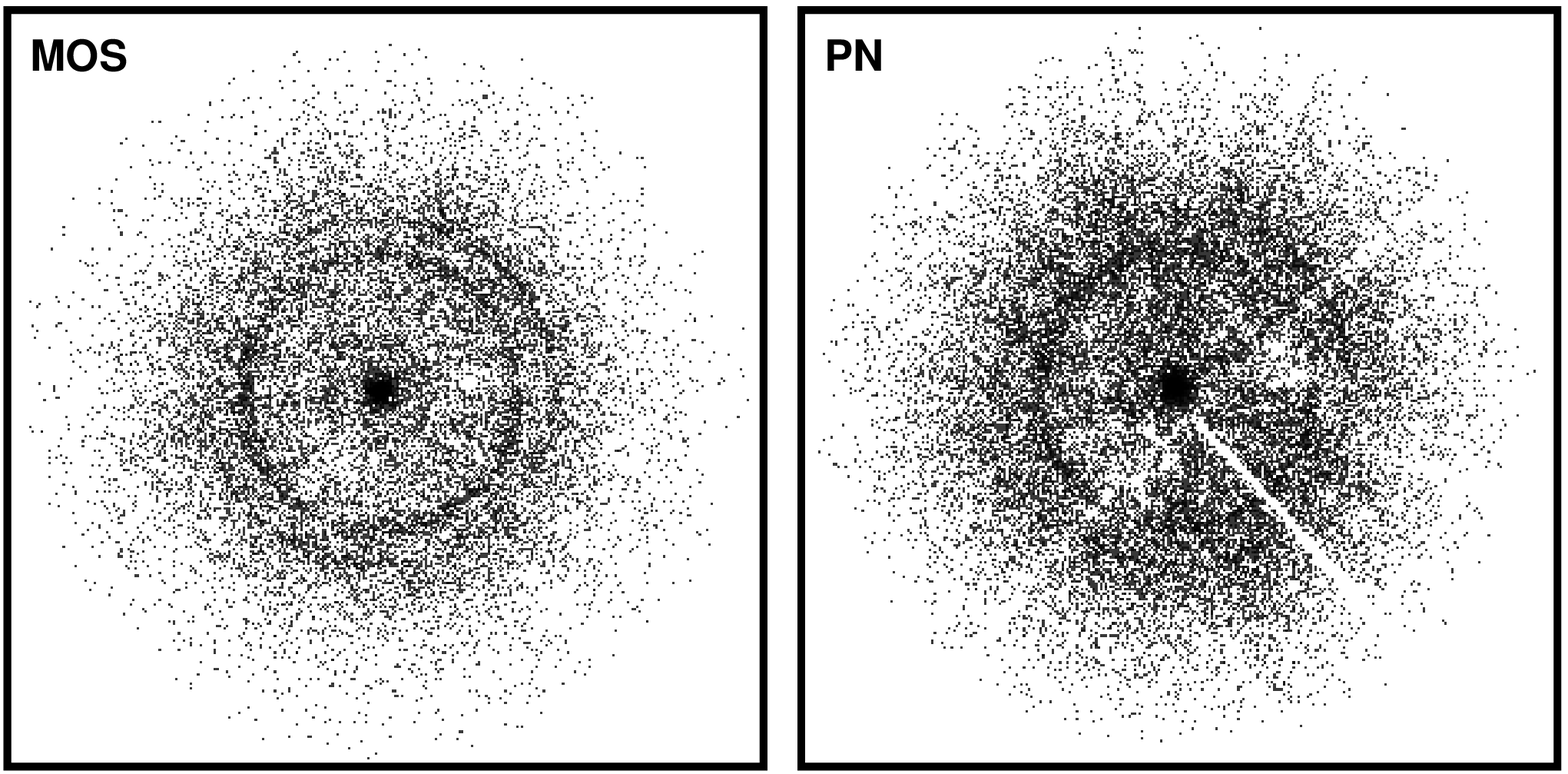}\hspace*{5mm}
            \includegraphics[height=50mm]{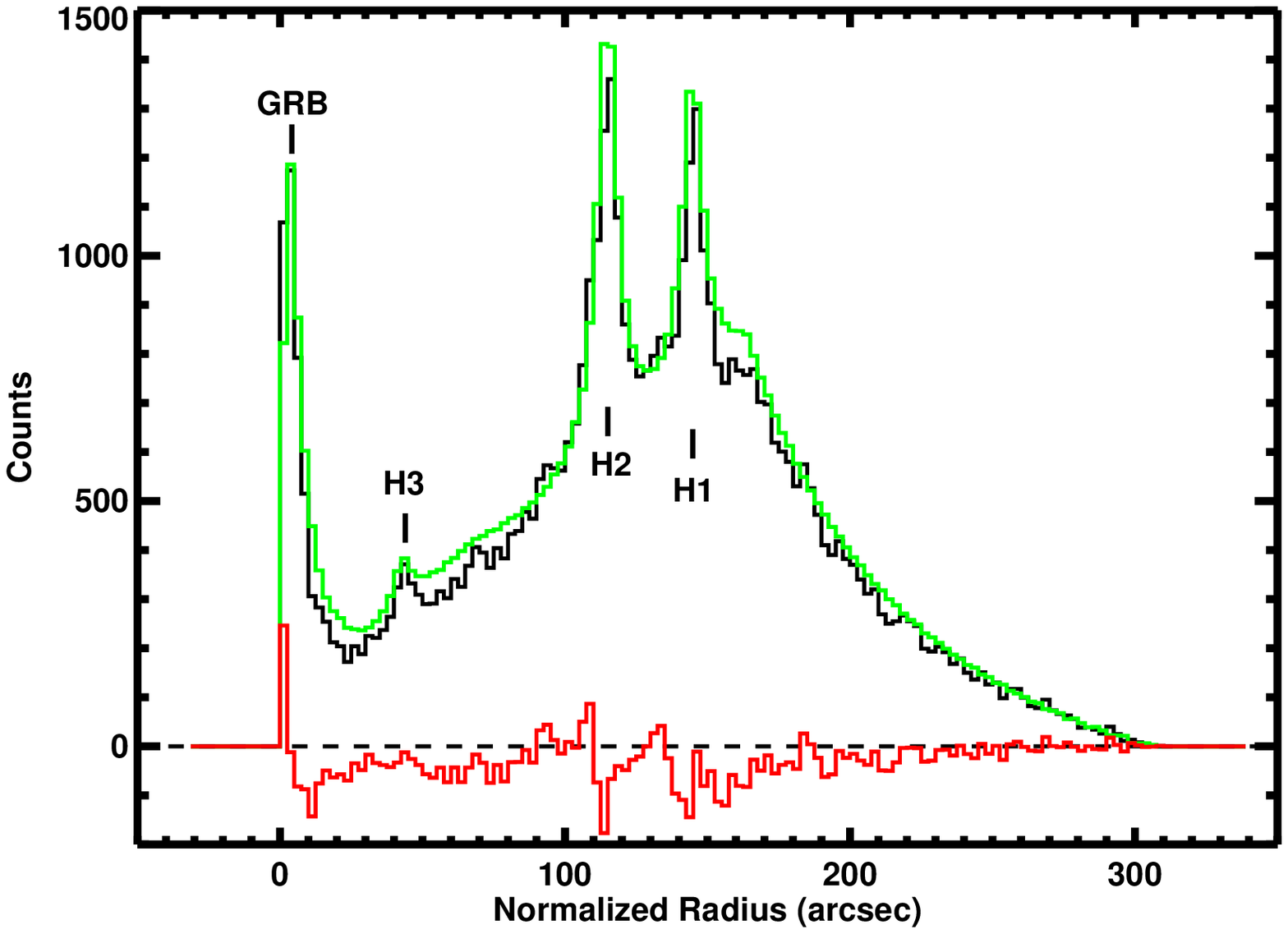}}
\caption[]{\normalsize
(left) Images of the 2-MOS and PN counts data in normalized
  coordinates ($x/\sqrt{t}$, $y/\sqrt{t}$).  The fading GRB afterglow,
  Halo~1 (outer), and Halo~2 (inner) are readily distinguished; the
  signature of the more distant and fainter Halo~3 (close to the GRB)
  is more subtle.  White streaks in the images are due to bright
  source exclusion regions and PN camera chip gaps, which are
  incorporated into our model via MOS and PN exposure maps.}
\label{fig:twodhisto}
\end{figure*}

We construct a 16-parameter, three-dimensional ($x$, $y$, $t$)
maximum-likelihood model including parameters to describe the fading
GRB afterglow, the expanding dust-scattered halos, and the vignetted
\xray\ (and charged particle-induced) background. The total counts
from all model components, for the 2-MOS and PN datasets, are
constrained to be equal to the observed count totals given above.  The
model probability density is evaluated at the position and arrival
time of each detected photon, the value of the probability density
function is interpreted as being proportional to the likelihood of
observing that event, and a global likelihood calculated; we then
attempt to maximize the global likelihood using numerical techniques,
specifically, the {\tt AMOEBA} minimization routine \citep{press} as
implemented in the IDL programming environment.  For ease of modeling,
we convert photon angular positions and arrival times as follows:
angular coordinates $x$ and $y$ are measured in arcsec relative to an
initial estimate of the GRB position, \xmm\ physical coordinates
(26024.8, 23674.1), with the $+x$ direction corresponding to
decreasing R.A. and $+y$ to increasing Dec., as usual.  The temporal
coordinate $t$ is measured relative to the burst trigger time ($t=0$),
scaled such that the \xmm\ observation begins at $t=1$, 22.2\,ksec
(6$^{\rm h}$10$^{\rm m}$44$^{\rm s}$) after the burst.  Using this
temporal coordinate, the \xmm\ observation extends from $t=1$ (defined
by the earliest arrival time in our dataset, from the MOS-2 detector)
to $t\approx 3.61$.  An illustration of the resulting dataset is
presented as an image in normalized coordinates ($x/\sqrt{t}$,
$y/\sqrt{t}$) in Fig.~\ref{fig:twodhisto}.

%
\begin{figure*}
\caption[]{\normalsize
(right) Histogram of 2-MOS + PN counts data (black), best-fit model
  (green), and residuals (lower panel), binned according to normalized
  radius, $r/\sqrt{t}$.  Contributions of the fading GRB afterglow
  (GRB) and the three expanding halos (H1, H2, H3) are individually
  distinguishable; the fall-off in counts at large normalized radius
  is due to our region of interest cut ($r_{\rm max}=311\arcsec$) as
  smeared out by the normalization procedure.}
\label{fig:modelhistogram}
\end{figure*}
%

We represent the fading GRB afterglow using an analytical model for
the \xmm\ Point Spread Function (PSF), which approximates the PSF as a
King function:
\beq
   s_p(r) = \frac{\alpha - 1}{\pi r_c^2}\; (1 + r^2/r_c^2)^{-\alpha},
\label{eq:psf}
\eeq
where $\alpha$ is the King slope, $r_{c}$ is the King core radius, $r$
is the radial distance from the source (ring centre), and the
integrated surface brightness is normalized to unity for $\alpha>1$.
Since we combine the data from the two MOS cameras, we use average MOS
PSF parameter values for the MOS data ($r_{c} = 4.916$ and $\alpha =
1.442$), and PN PSF values for the PN data\footnote{PSF model and
  parameter values are obtained from the \xmm\ calibration document
  ``In Flight Calibration of the PSF for the PN Camera'' by Simona
  Ghizzardi, located at
  http://xmm.vilspa.esa.es/docs/documents/CAL-TN-0029-1-0.ps.gz.  See,
  in particular, Eq.\,1 and Table\,1 in the document.} ($r_{c} =
6.636$ and $\alpha = 1.525$).  The fading of the afterglow is
parameterized as a power-law with temporal index $\alpha_g$,
referenced to the burst trigger time.  

We represent the expanding, fading dust-scattered halos, centred on
the GRB position, using a numerical model for ring-like sources
observed with \xmm\ that we have developed.  The derivation of this
model is provided in Appendix~\ref{app:rings}; as realized in our
code, this bivariate function is precalculated on a grid and
interpolated to the particular values required for each function
evaluation.  Fading of the halos is parameterized as a power-law with
temporal index $\alpha_h$, referenced to the burst trigger time; this
index is assumed to be the same for all rings.  In order to avoid
complications associated with normalizing the model probability
density function for indices $\alpha_h\le -1$ as well as for
$\alpha_h>-1$, we impose a hard limit $\alpha_h\ge -0.95$ on this
parameter during fitting.  Initial investigations suggested that this
limit was encountered in less than 5\% of bootstrap trials; however,
our final analysis resulted in a somewhat greater proportion of 9\% of
all trials encountering this limit.

Because the rings expand by a factor of $\sqrt{3.6}\approx 1.9$ over
the course of the observation, and because we wish to accurately model
the background over the region of interest, we find it necessary to
apply an off-axis vignetting function \citep{lfs+03} to the rings and
background:  
\begin{equation}
  S(r) = S(0)(1 - mr^e)
\end{equation}
where $S(0)$ is the surface brightness on-axis, $S(r)$ is the surface
brightness at off-axis distance $r$ (measured in arcsec), and $m$ and
$e$ are the vignetting parameters for the EPIC cameras.  We use $m =
3.43 \times 10^{-5}$ and $e = 1.45$ for the MOS cameras, and $m = 3.51
\times 10^{-5}$ and $e = 1.53$ for the PN, which we find accurately
models the vignetting of PN and MOS exposure maps generated in SAS.  

The excluded sources and the PN exposure map present discontinuities
to the model probability density function, which must be normalized to
yield an appropriate fit.  We perform this normalization via numerical
integration over a three-dimensional rectangular grid ($x$, $y$, $t$),
where the grid spacing is 2\arcsec\ in $x$ and $y$, and the temporal
grid is divided into 40 equal intervals.  The same cuts applied to the
data, including the circular region of interest, the exclusion regions
around identified bright sources, and (for PN only) the PN exposure
map, are applied to the grid prior to integration, and are thus
(approximately) accounted for.

We maximize the likelihood function using the {\tt AMOEBA} algorithm,
in stages, using multiple fitting rounds at each stage (with fit
starting points randomized prior to each round) to ensure a relatively
wide-ranging exploration of the parameter space.  In sequence, we
execute: (1) Ten rounds of fitting the GRB counts parameters and
fading index $\alpha_g$; (2) Ten rounds of fitting with all ring
parameters (counts, radii, and fading index $\alpha_{r}$) also free;
and finally, (3) Five rounds of fitting with all parameters free.

Initial examination of the data suggested the presence of a third
dust-scattered halo at small radius
(Fig.~\ref{fig:twodhisto}). Preliminary investigation confirmed the
reality of this feature, and indicated that it should be interpreted
as due to \xray\ scattering off a third dust sheet, more distant than
the two discussed by previous authors (\S\ref{sub:discuss:halo}).
We therefore incorporate this third halo into our fits, with the same
burst time $t_0$ and fading index $\alpha_h$ as the others.

Uncertainties in our model parameters are derived with the bootstrap
Monte Carlo method, executing multiple trials and accumulating
best-fit parameter values from each trial to realize the posterior
distribution of our model parameters.  For each trial, a new dataset
is generated by drawing randomly, with replacement, from the true
dataset until we have the same number of events as in the actual data.
The fitting sequence, described above, is identical for each trial;
parameter starting points are randomized based on preliminary
estimates (subsequently verified in our final analysis) of the
uncertainties in each parameter.  The result should approximate the
posterior distributions that would result from a full Bayesian
analysis using uniform (non-informative) prior distributions for each
of the model parameters.  For our final analysis, presented below, we
executed 1000 trials in all.

The optimization of nonlinear functions in many-dimensional spaces is
a known hard (``NP'') computational problem.  We therefore point out
that the bootstrap Monte Carlo approach is expected to yield
conservative estimates of parameter uncertainties in cases where the
model fails to offer a statistically-complete representation of the
data, or where the underlying fitting procedure fails to identify the
global minimum in each and every trial.  The reason is that, quite
simply, the failure to identify the best possible fit is robustly
expected to drive the fitting routine to explore a larger (rather than
smaller) region of parameter space.  This is true as long as the
fitting routine is not trapped in a single local minimum, repeatedly,
in trial after trial; in our analysis we avoid this numerical disorder
by randomizing parameter starting values for each trial, as mentioned
above.


\subsection{Results}
\label{sub:obs:results}

The results of our model fitting, including parameter uncertainties
quoted both as standard deviations and as 90\%-confidence intervals,
are presented in Table~\ref{tab:param}.  90\%-confidence intervals are
defined as the minimum-length intervals that encompass 90\% of the
parameter values from our bootstrap Monte Carlo trials (e.g., 900 of
1000 parameter values).  Associated background count totals can be
derived by subtracting the counts of the model components given in the
table from the total instrument counts of 21,289 for 2-MOS and 32,085
for PN (\S\ref{sub:obs:model}).


\begin{table*}
\begin{minipage}{96mm}

\caption{Fit parameters for GRB\,031203 afterglow and expanding
  halos}

\label{tab:param}

\begin{tabular}{lrrrl}

\hline

Parameter & Median & $\sigma$ & \multicolumn{2}{c}{90\%-conf.} \\

\hline

GRB $x$ & 0.46 & 0.14 & 0.23 & 0.70 \\
 \phm{GRB} $y$ & 0.19 & 0.13 & $-0.04$ & 0.40 \\
 \phm{GRB} 2-MOS counts\hspace*{3em} 
         & \hspace*{0.5em} 2252.7 
         & 60.3
         & \hspace*{0.7em} 2157.3 & 2355.6\hspace*{0.7em} \\
 \phm{GRB} PN counts & 2340.0 & 125.5 & 2172.4 & 2559.2 \\
 \phm{GRB} Fading index, $\alpha_g$ 
         & $-0.557$ & 0.056 & $-0.64$ & $-0.46$ \\

Halos $t_{0}$ (s) 
         & 11.0 & 416.9 & $-620.6$ & 722.8 \\
 \phm{Halos} Fading index, $\alpha_h$
         & $-0.77$ & 0.10 & $(-0.95)$ & $-0.65$ \\

Halo 1 Radius (arcsec) & 145.04 & 0.70 & 143.74 & 146.07 \\
 \phm{Halo 1} {\it Distance} (pc) & 871.7 & 9.7 & 855.6 & 886.8 \\
 \phm{Halo 1} 2-MOS Counts & 1365.0 & 67.8 & 1251.1 & 1467.4 \\
 \phm{Halo 1} PN Counts & 1686.4 & 88.8 & 1568.4 & 1855.6 \\

Halo 2 Radius (arcsec) & 114.93 & 0.63 & 113.90 & 115.94 \\
 \phm{Halo 2} {\it Distance} (pc) & 1388. & 14. & 1364. & 1410. \\
 \phm{Halo 2} 2-MOS Counts & 2059.3 & 65.5 & 1952.9 & 2164.1 \\
 \phm{Halo 2} PN Counts & 2229.3 & 108.9 & 2041.6 & 2389.5 \\

Halo 3 Radius (arcsec) & 42.97 & 0.81 & 41.46 & 44.06 \\
 \phm{Halo 3} {\it Distance} (kpc) & 9.94 & 0.39 & 9.32 & 10.52 \\
 \phm{Halo 3} 2-MOS Counts & 268.5 & 40.7 & 203.4 & 334.1 \\
 \phm{Halo 3} PN Counts & 219.7 & 32.2 & 177.7 & 274.8 \\


\hline

\end{tabular}

\medskip

Median, standard deviation about the median ($\sigma$), and
90\%-confidence intervals are derived from model fits to 1000
bootstrap Monte Carlo realizations of the dataset.  PN and 2-MOS
counts are defined as the total counts gathered from the PN (0.2 to
4.0\,keV) and two MOS cameras (0.2 to 3.0\,keV), respectively.  Halo
radii are defined as the radius in arcsec at the start of
\xmm\ observations.  Halo distances are determined by the best-fit
halo radii and $t_0$ values from each trial and are not independent
parameters.  GRB coordinates ($x$, $y$) are offsets in arcsec from
\xmm\ physical coordinates (26024.8, 23674.1); the median offset of
(0.46, 0.19) arcsec corresponds to R.A. 08$^{\rm h}$02$^{\rm m}$30\fs
20, Dec.\ $-$39\arcdeg 51\arcmin 02\fs 8 (J2000) using the uncorrected
\xmm\ aspect. $t_{0}$ is the time of zero-radius for the expanding
halos relative to the \integral\ trigger time of 22:01:28 UT.
The model enforces a hard limit $\alpha_h\ge -0.95$ which is
encountered in 9\% of our bootstrap trials.

\end{minipage}
\end{table*}


To illustrate the model goodness-of-fit, we present in
Fig.~\ref{fig:modelhistogram} a histogram of the data (2-MOS + PN
counts) compared to our best-fit model.
%
%
%
%
In this figure, as well as in
Fig.~\ref{fig:twodhisto}, we present the data in terms of the
``normalized radius,'' which is the radial distance from the GRB
afterglow position, divided by the square root of the normalized time
(time measured from the GRB trigger, with $t=1$ at the start of the
\xmm\ observation).  This normalization serves to highlight the
expanding halos as distinct features, while not significantly altering
the appearance of the afterglow itself (at small radius); the fall-off
in counts at large radius is due to our region of interest cut
($r_{\rm max}=311\arcsec$), as smeared out by the conversion to
normalized units.  We find the overall fit to be satisfactory.
Specifically, although several systematic features are apparent in the
residuals, we have investigated these without identifying any obvious
model failings that might account for them.  In particular, we have
verified that none of the features are due to unaccounted-for bright
point sources or instrument artifacts (e.g., hot pixels or hot
columns).  Variations in effective exposure within our region of
interest, not accounted for by our simple vignetting model, may be
responsible for some of these trends.  Note that our fitting procedure
is executed against the unbinned data, in three dimensions, and that
this figure merely represents one possible projection of the model and
data onto a one-dimensional space, for visualization purposes.

In Fig.~\ref{fig:multi} we present the full posterior distribution for
four parameters of special interest: $\alpha_g$, the temporal
power-law index for the fading of the GRB afterglow; $t_0$, the time
of zero-radius for the expanding halos; $\alpha_h$, the temporal
power-law index for the fading of the expanding halos; and the sum of
the 2-MOS and PN counts for Halo~3, the halo due to the most distant
scattering dust sheet, discovered in our analysis.  90\%-confidence
intervals on each of these parameters are indicated.

\section{Discussion}
\label{sec:discuss}

In the sections below we discuss the prior understanding of the
\grbone\ \xray\ blast phenomenon and our own results as to the
properties and timing of the event.  We then discuss the full set of
observations of \grbone\ in the context of the known varieties of
cosmic explosion, including \grbtwo.


\subsection{Previous Findings}
\label{sub:discuss:previous}

\citeta{vwo+04} originally reported the discovery of two evolving
dust-scattered \xray\ halos centred on \grbone\ and made the first
attempt to model the halos and extract parameters of the \xray\ blast
and associated dust scattering sheets along the line of sight within
the Milky Way.  They constrained the time of burst for the inner and
outer rings to be $t_{0} = 2794^{+2765}_{-3178}$\,s and $t_{0} =
2005^{+2512}_{-2867}$\,s respectively, relative to the
\integral\ trigger, and by applying a model of the dust properties,
determined that the initial soft \xray\ event had a power law spectrum
with photon index $\Gamma\approx 2$.

Subsequently, \citeta{whl+04}, \citeta{wvw+06}, and \citeta{tm06},
using different analysis approaches, refined the details of this
picture.  They derived more precise distances to the dust sheets
($d_1\approx 870$\,pc and $d_2\approx 1390$\,pc, respectively), and
correspondingly more precise constraints on the timing of the
\xray\ blast.  The time constraint from \citeta{wvw+06} is $t_{0} =
600 \pm 700$\,s at 90\% confidence; the inferred power-law spectral
index for the \xray\ blast is reported as either $\Gamma=2.5\pm 0.3$
\citepa{whl+04} or $\Gamma=2.1\pm 0.2$ \citepa{tm06}, results which
are consistent at the $\approx$1$\sigma$ level.  The two conclusions
as to the fluence of the \xray\ blast over 1--2 keV, $F_{\rm X} =
1.5(3)\times 10^{-6}$\,\ergcmsq\ \citepa{wvw+06} versus $3.6(2)\times
10^{-7}$\,\ergcmsq\ \citepa{tm06}, are more discrepant, which
\citeta{tm06} attribute to the authors' diverging assumptions on the
relation between \xray\ scattering optical depth and extinction.

In discussing their model for the dust properties, \citeta{wvw+06}
show a plot of the halo fading (their Fig.~2) which corresponds to a
temporal power-law index of $\alpha_h\approx -0.6$.  They use this
fact (along with the size of the scattering halos) to derive a maximum
size for the dust grains, $a_{\rm max} = 0.50\pm 0.03$\,\um; in
accounting for the effects of smaller grains they use a power-law size
distribution with power-law index equal to $-3.5$.


\subsection{Discovery of a Third Halo}
\label{sub:discuss:halo}

\begin{figure*}
\centerline{\includegraphics[height=60mm]{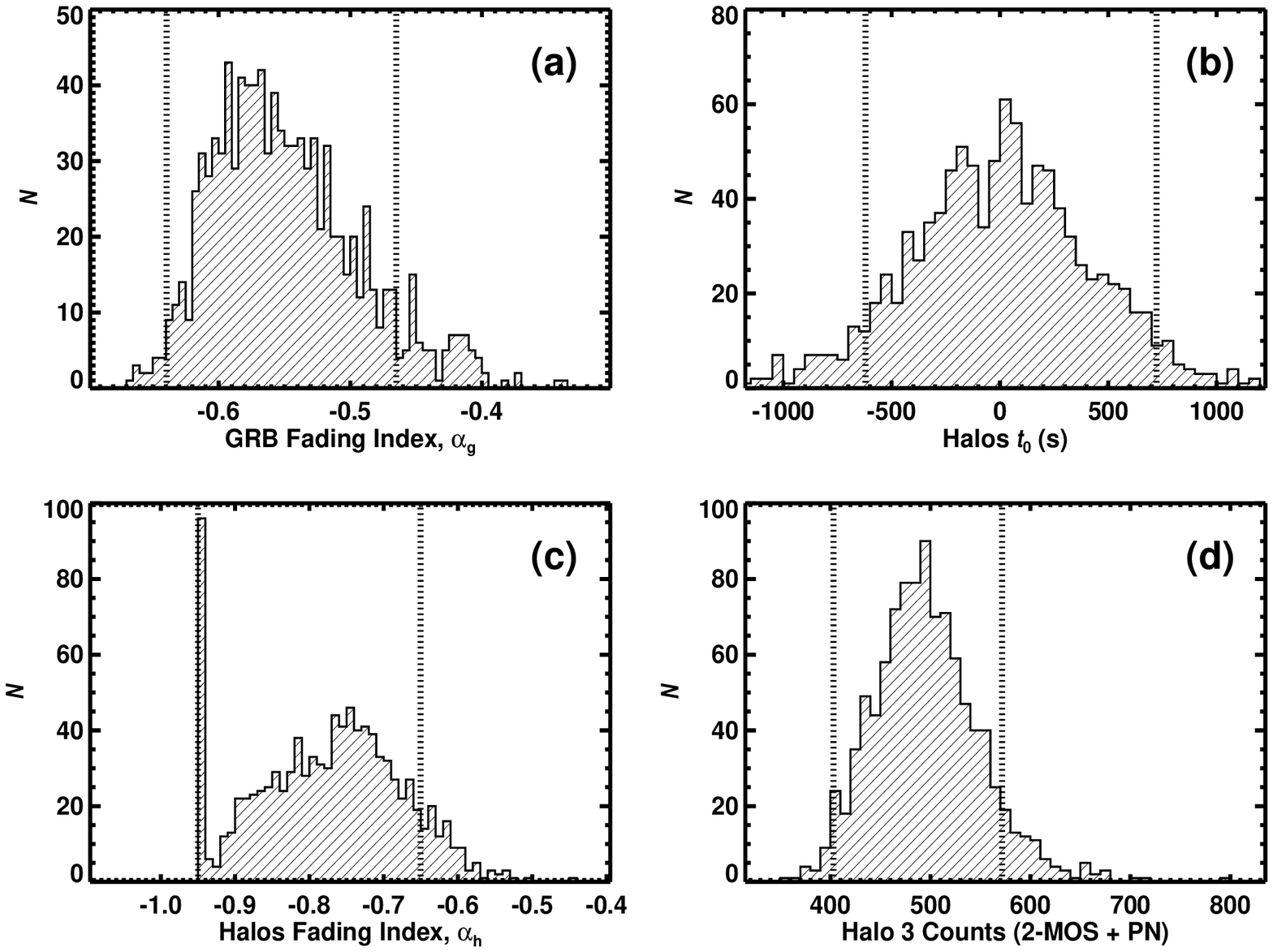}\hspace{5mm}%
            \includegraphics[height=60mm]{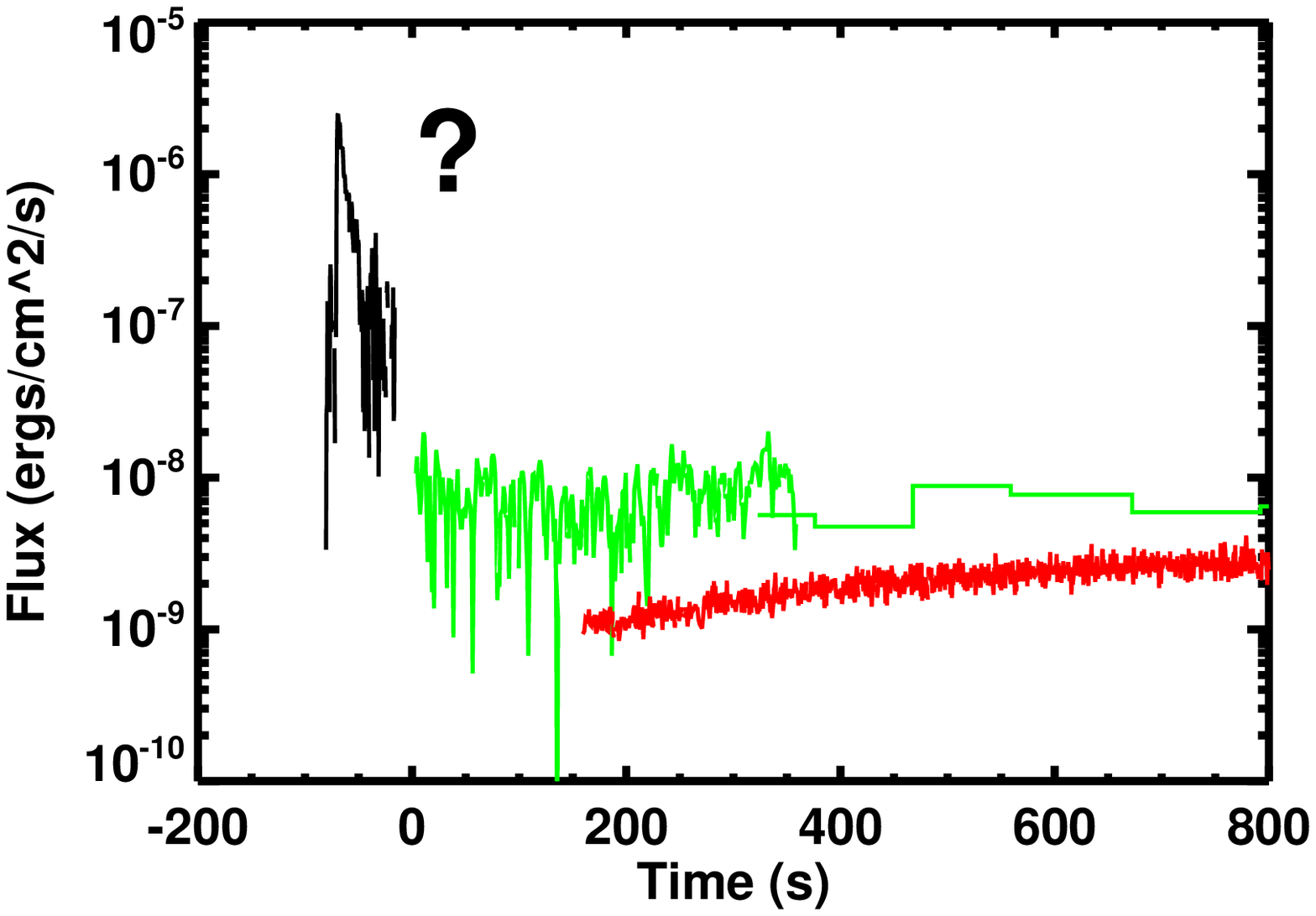}}
\caption[]{\normalsize
(left) Approximate posterior distributions of selected model
  parameters, as derived from our bootstrap Monte Carlo analysis.  (a)
  The GRB afterglow temporal fading index, $\alpha_g$; (b) $t_0$ for
  the expanding halos, corresponding to the characteristic emission
  time of the \xray\ blast, measured relative to the trigger time for
  \grbone; (c) The halos temporal fading index, $\alpha_h$ (note this
  parameter is subject to a hard limit, $\alpha_h\ge -0.95$); (d) The
  sum of 2-MOS + PN counts for the expanding Halo~3, discovered in our
  analysis.  90\%-confidence intervals for each parameter are
  indicated graphically (dotted lines).}
\label{fig:multi}
\end{figure*}

Our analysis reveals the presence of a third, faint halo
(Figs.~\ref{fig:twodhisto}, \ref{fig:modelhistogram}) in addition to
the two identified by previous authors.  The presence of this halo is
robust in our analysis, with associated total 2-MOS + PN counts
between 403 and 572 at 90\%-confidence (Fig.~\ref{fig:multi}d), and
with roughly equal counts in the two sets of detectors, which are fit
independently (Table~\ref{tab:param}).

Initially, we speculated that the third halo might be the signal of a
second \xray\ blast or early afterglow emission, with the
\mbox{X-rays} scattering off one of the two previously known dust
sheets.  If this were the case, the third halo would necessarily be
associated with a distinct (later) $t_0$ and would share the same
distance $d$ as one of the other halos.

To investigate this possibility, we performed a distinct bootstrap
Monte Carlo analysis of 600~trials, allowing $t_0$ for the third halo
to vary while fixing $t_0=0$ for the two other halos.  Given the
angular size of the halo at the start of the \xmm\ observation,
$\theta$, $t_0$ is directly related to the distance to the dust
scattering sheet by $d=2c(t-t_0)/\theta^2$, where $t-t_0$ is the time
from the burst to the start of \xmm\ observations.  This analysis
yields a 90\%-confidence interval for $d_3$ of 8597 to 10503\,pc,
inconsistent with both $d_1\approx 870$\,pc and $d_2\approx 1390$\,pc.
Hence, we conclude that the third halo is due to a third dust sheet
and require $t_0$ to be identical, for all halos, in our full
analysis.

The third dust sheet is located at greater distance and greater
Galactic radius ($R\approx 14.5$\,kpc) than the two nearby sheets, and
nominally at greater distance from the Galactic plane as well
($z\approx -0.83$\,kpc), given the Galactic coordinates toward
GRB\,031203, $l=255.7\arcdeg$ and $b=-4.8\arcdeg$.  However,
reconstructions of the three-dimensional distribution of neutral
hydrogen density throughout the Milky Way \citep{kdk+07,kd08} reveal a
``warp'' in the disk which has an amplitude of $\delta z\approx
1$\,kpc at $R=15$\,kpc, in precisely the ($-z$) sense needed to place
the third dust sheet within or near the densest portion of the
\HI\ disk.  The existence and location of the third dust sheet is not
particularly surprising in this context.

%
\begin{figure*}
\caption[]{\normalsize
(right) Superposition of \grbtwo\ \swift\ BAT (green, 15--350\,keV)
  and XRT (red, 0.3--10 keV) light curves with a hypothetical
  precursor event having properties identical to \grbone\ (black,
  20--200 keV).  The start of \swift\ BAT observations defines $t=0$;
  the \integral\ light curve of \grbone\ is obtained from
  \citeta{sls04}, rebinned, corrected for redshift and time dilation,
  and offset by $-70$\,s to represent its occurrence prior to the
  start of \swift\ observations.  Negative (background-subtracted)
  flux measurements are not shown on this logarithmic plot.}
\label{fig:swiftintlc}
\end{figure*}
%

Our final analysis yields 90\%-confidence intervals on the distances
to the dust sheets as follows: $d_1 = 856$--887\,pc, $d_2 =
1364$--1410\,pc, and $d_3 = 9321$--10523\,pc (Table~\ref{tab:param});
these uncertainties are inclusive of uncertainties in the value of
$t_0$ itself, which we have not fixed (e.g., Fig.~\ref{fig:multi}b).
For Halo~1 and Halo~2, our values agree with those derived by
\citeta{wvw+06} ($d_1 = 868_{-16}^{+17}$\,pc and $d_2 =
1395_{-30}^{+15}$\,pc) and \citeta{tm06} ($d_1 = 870 \pm 5$\,pc and
$d_2 = 1384 \pm 9$\,pc); note, however, that \citeta{tm06} fix
$t_0=0$\,s in their analysis.

The median value of the halo fading index from our analysis is
$\alpha_h = -0.77$ with a 90\%-confidence interval $\alpha_h < -0.65$
that is unconstrained from below because of our hard limit
($\alpha_h\ge -0.95$); Fig.~\ref{fig:multi}c presents the full
posterior distribution.  Our result appears mildly in conflict with
the value $\alpha_h\approx -0.6$ from \citeta{wvw+06}, with associated
implications for the properties of the scattering dust particles (see
also \citeta{tm06}); we have not investigated these implications in
detail.


\subsection{Properties of the \grbone\ \xray\ Blast}
\label{sub:discuss:timing}

Our analysis yields a 90\%-confidence interval on the timing of the
\xray\ blast which places it within roughly ten minutes of the
\grbone\ trigger, between $-621$\,s and $+723$\,s, with median value
and standard deviation $t_0 = 11\pm 417$\,s; the full posterior
distribution for $t_0$ is shown in Fig.~\ref{fig:multi}b.

Our constraint is comparable in precision to the best previous
determination, $t_0 = 600 \pm 700$\,s at 90\%-confidence
\citepa{wvw+06}.  The shift of our confidence interval to earlier
times sharpens the contrast between the implications of the \xmm\ and
\integral\ observations: The \integral\ pointing extends to 300\,s
after the GRB trigger, and only 22.5\% of our trials yield $t_0$
values consistent with this constraint.  If we assume, in accordance
with standard GRB models, that the \xray\ blast could not have
preceded the \gray\ trigger, then our upper limit of $t_0 < 545$\,s at
90\%-confidence is also relevant in this context.

An alternative approach to resolving this dilemma would reduce the
peak flux of the \xray\ blast -- making it undetectable in the
\integral\ observation -- by extending its fluence over a significant
time interval after the GRB.  This hypothesis is also constrained by
our analysis, since we observe no evidence that the halos are radially
resolved; our analytical model for unresolved ring-like sources
(Appendix~\ref{app:rings}) provides a satisfactory fit to
the radial profiles of the halos (Fig.~\ref{fig:modelhistogram}). 

While we have not constrained the duration of the \xray\ blast
directly, our results on the timing of the blast demonstrate that our
model is sensitive to changes of $\delta t_0\approx 600$\,s, and we
expect that, in demonstrating an adequate fit with the unresolved
``ring'' model, we have constrained the duration of the blast at
approximately this magnitude, $\delta t\simlt 600$\,s.  Given that the
\xmm\ observation began $t=22.2$\,ksec after the GRB trigger, a
duration of 600\,s corresponds to $\delta t/t=3\%$ or $\delta
r_1=4\arcsec$ for Halo~1, which would be most sensitive to these
effects.  Since the \xmm\ PSF has a width of ${\rm FWHM}=4\arcsec$, it
is reasonable to think the data would be sensitive to signs of
broadening at this level.  Moreover, the dust sheets themselves must
be extended to some degree; any such extension would tend to further
broaden the radial profile of the halos, beyond any broadening due to
the finite duration of the blast itself.

Finally, we note that \citeta{whl+04} call attention to the relatively
slow fading of the \xray\ afterglow of \grbone\ during the
\xmm\ observation, $\alpha_g\approx -0.55$ (Table~\ref{tab:param}),
which was significantly flatter than the canonical $\alpha_{\rm
  X}\approx -1.3$ familiar from \bepposax\ observations at the time. In
retrospect we can identify this as one of the first observations of a
``plateau phase'' in GRB \xray\ afterglows, now familiar from
\swift\ observations \citep[e.g.,][]{nkg+06,rlb+09}.  As we have since
learned, extrapolation of this slow decay to early times, $t\simlt
1000$\,s, is not particularly revealing of the likely \xray\ flux at
that time.


\subsection{Could \grbone\ and \grbtwo\ be Cosmic Twins?}
\label{sub:discuss:twins}

As mentioned in \S\ref{sec:intro}, a GRB similar to \grbone, as
observed by \integral, could have occurred immediately prior to
\swift\ observations of \grbtwo\ without being observed by
\swift. Likewise, a slow-evolving XRF similar to \grbtwo\ could have
occurred following \grbone\ without being observed by \integral, if it
exhibited either a sufficiently low peak flux or peaked after
$t=300$\,s, when \integral\ began its slew to a new pointing.  In this
scenario, then, both events would consist of two separate bursts, one
peaking at \gray\ energies and one peaking in the \xray.  Importantly,
while we have evidence for distinct \gray\ and \xray\ emission
episodes in \grbone, in this picture only the \xray\ episode was
observed from \grbtwo.  

A variation on this scenario, without any separate episode of
\gray\ emission from \grbtwo, was proposed by \citet{ggm+06}.  These
authors were concerned with the question of whether the prompt
high-energy emissions of \grbone\ and \sngrbb\ could be reconciled
with the high-energy relationships between peak energy and
isotropic-equivalent energy output that are observed among the
$z\simgt 1$ GRB population.  They concluded, as we do here, that the
observation of \grbtwo\ by \swift\ suggests strongly that a similar
phenomenon might be responsible for the ``\xray\ blast'' inferred from
the expanding \xray\ halos of \grbone. 

To illustrate the nature of our proposal, we have generated a figure
that superimposes the prompt-emission light curve of
\grbone\ \citepa{sls04}, appropriately rescaled to the redshift and
luminosity distance of \grbtwo, on the \swift\ BAT and XRT lightcurves
of the prompt and early emission of
\grbtwo\ (Fig.~\ref{fig:swiftintlc}).  
%
%
%
%
With $t=0$ at the beginning of \swift\ observations, it is necessary
to offset the \integral\ light curve to earlier times in order to
avoid detection by the \swift\ BAT; we arbitrarily choose a $-70$\,s
offset.  The BAT light curve is a combination of the mask-weighted
light curve generated by the BAT standard analysis ($t<350$\,s) with
data from \citet{cmb+06} (64-s integrations at $t>350$\,s).  The XRT
light curve (red) for \grbtwo\ is obtained from the \swift/XRT light
curves repository at the University of Leicester\footnote{\swift/XRT
  light curves repository: \url{http://www.swift.ac.uk/xrt\_curves/}}
\citep{ebp+07,ebp+09}.

As Fig.~\ref{fig:swiftintlc} shows, this ``cosmic twin'' to
\grbone\ would have exhibited a peak flux of
$2.7\tee{-6}$\,\ergcms\ (20--200 keV), roughly 11 times brighter than
the original event, which occurred at about three times the distance of
\grbtwo.  This addresses an important condition on our hypothesis,
which is that the burst not be bright enough to trigger detectors of
the Interplanetary Network \citep[IPN; e.g.,][]{hmk+06}, and in
particular, \konuswind\ \citep{afg+95}, which did not report any
trigger associated with \grbtwo.  A review of the catalog of
\konuswind\ short burst detections\footnote{\konuswind\ short burst
  catalog: \url{http://www.ioffe.ru/LEA/shortGRBs/Catalog/}}
reveals that the peak flux distribution peaks at
$7.5\tee{-6}$\,\ergcms\ (15--2000 keV), although bursts are detected
down to a minimum peak flux of $2\tee{-6}$\,\ergcms.  Depending on the
high-energy spectrum of \grbone, the peak flux of its hypothetical
$z=0.033$ cosmic twin would be from $5.2\tee{-6}$ to
$7.9\tee{-6}$\,\ergcms\ (15--2000 keV), likely below the regime of
maximum sensitivity for this experiment.  With a peak flux falling
close to threshold in the brightest case, relatively slight variations
in the profile or spectrum of the event would be sufficient to render
the burst undetectable.

An associated ``classical GRB'' like \grbone\ thus seems plausible;
however, our analysis indicates that the \xray\ properties of the two
events must differ to some degree. First, we have constrained the
timing of the \grbone\ \xray\ blast to fall within $t_0=11\pm 417$\,s
of the GRB ($t_0 < 545$\,s at 90\%-confidence;
\S\ref{sub:discuss:timing}), whereas the XRT lightcurve for
\grbtwo\ rises slowly until it peaks at $t\approx 1000$\,s.  Second,
depending on the assumed relation between extinction and
\xray\ scattering optical depth \citepa{tm06}, the
\grbone\ \xray\ blast may have released as much as three times the
energy of the prompt \xray\ emission of
\grbtwo\ (Table~\ref{tab:comp}).

To illustrate these constraints in detail, and explore other possible
explanations for the \grbone\ \xray\ blast, Table~\ref{tab:comp}
presents the \gray\ and \xray\ properties of all five confirmed
spectroscopic GRB-SNe, and the properties of two other cosmic
explosions that may be interesting in this context, \grbflare\ and
\snxrt.


\begin{table*}
\begin{minipage}{128mm}

\caption{Properties of GRB\,031203 and related cosmic
  explosions} 

\label{tab:comp}

\begin{tabular}{rlcrrrrc}

\hline

\multicolumn{2}{c}{Event} &
$z$ &
$\log E_{\gamma{\rm ,iso}}$ &
$\log E_{\rm X,iso}$ &
$\Delta t_{\text{X} - \gamma}$ & 
Interpretation &
Refs. \\ \hline

SN  & 1998bw  &0.0085 &   47.8 & 47.1 &   +10   & GRB-SN\,Ic & 1,2 \\
XRF & 020903  & 0.251 &   48.5 & 48.9 &     0   & XRF-SN\,Ic & 3,4 \\
GRB & 030329  & 0.167 &   52.0 & 51.1 &    +3   & GRB-SN\,Ic & 5 \\
GRB & 031203 & 0.105 & 50.0 & 49.2--49.6 &$\pm$600 & GRB-SN\,Ic + ? & 6,7,8,9 \\ 
GRB & 050502B & (2.0) &   51.4 & 51.0 &  +250   & GRB + \xray\ flare & 10,11 \\
XRF & 060218  & 0.033 &   49.4 & 49.1 & +1050   & XRF-SN\,Ic + shock & 12 \\
SN  & 2008D   &0.0065 &$<$46.5 & 46.3 &    --   & SN\,Ic shock & 13 \\

\hline

\end{tabular}

\medskip

$\log E_{\gamma{\rm ,iso}}$ is the isotropic-equivalent gamma-ray
energy over 20 keV to 2 MeV, rest-frame (ergs); $\log E_{\rm X,iso}$
is the isotropic-equivalent prompt \xray\ energy over 2 to 10 keV,
rest-frame (ergs); and $\Delta t_{\text{X} - \gamma}$ is the mean or
estimated delay between gamma-ray and \xray\ emission, rest-frame (s).
A redshift of $z=2$ is assumed for GRB\,050502B; no redshift-related
corrections are applied for SN\,1998bw and SN\,2008D. The two values
of $\log E_{\rm X,iso}$ for \grbone\ are from \citeta{tm06} (49.2) and
\citeta{whl+04} (49.6), respectively.  The \xray\ energy and time
delay for \grbtwo\ are derived by analysis of the XRT lightcurve from
the \swift\ XRT lightcurves repository \citep{ebp+07,ebp+09}.  The
upper limit on $E_{\gamma{\rm,iso}}$ for SN\,2008D is extrapolated
from the \swift\ BAT upper limit using an $\alpha=1.5$ power-law
spectrum.   
References:  
 $^1$\citet{gvv+98}; $^2$\citet{fac+00};
 $^3$\citet{skb+04a}; $^4$\citet{slg+04}; 
 $^5$\citet{vsb+04};
 $^6$\citeta{sls04}; $^7$\citeta{whl+04}; $^8$\citeta{tm06}; $^9$this work; 
 $^{10}$\citet{brf+05};  $^{11}$\citet{sbb+08};
 $^{12}$\citet{cmb+06}; 
 $^{13}$\citet{sbp+08}.

\end{minipage}
\end{table*}


First, we consider the ``cosmic twins'' hypothesis, comparing the
properties of \grbone\ and \grbtwo. A distinct, earlier episode of
high-energy emission for \grbtwo\ is readily accommodated by the total
\gray\ energy budget of \grbone; in order to avoid detection by
various IPN experiments, the peak luminosity, peak energy, or total
energy release should be somewhat reduced from \grbone\ values.  The
\xray\ blast from \grbone\ released more energy in the \xray\ than
\grbtwo; however, depending on the assumed relation between extinction
and \xray\ scattering optical depth, the additional 25\% \citepa{tm06}
to 200\% \citepa{wvw+06} energy requirement is not necessarily severe.
Finally, the time delay from burst trigger to \xray\ peak
(technically, to the characteristic time of emission) should be at
least twice as long for \grbtwo\ as for \grbone.  None of these
requirements appear to us prohibitive; collectively, however, they
restrict the allowed parameter range of models considerably.  The main
appeal to satisfying these constraints, in an Occam's Razor sense, is
the requirement for one fewer \textit{sui generis} type of GRB-SN.

Next, we consider the \xray\ flare explanation for the
\grbone\ \xray\ blast \citepa{wvw+06}.  Even at the maximum inferred
energy of $4\tee{49}$\,erg (2--10\,keV), the ratio of \xray\ to
\gray\ energies for \grbone\ would be no greater than observed in
\grbflare.  Moreover, the relative timing of the \gray\ and
\xray\ episodes is very close to what is needed to satisfy both our
time constraint, derived from the expanding halos, and the
\integral\ slew time constraint, $t>300$\,s.  The latter applies to
any relatively short (hence high peak flux) model for the
\xray\ blast.  At the same time, the \xray\ flare from
\grbflare\ remains an extreme example of its class
\citep[e.g.,][]{cmr+07}, and in this model the \grbone\ \xray\ blast
would be a similarly extreme case; in addition, it would be the only
known example of an \xray\ flare from a GRB-SN.

None of the other cosmic explosions listed in Table~\ref{tab:comp} are
known to exhibit strong distinct episodes of \gray\ and
\xray\ emission.  Their properties are presented to illustrate the
diversity of the GRB-SN phenomenon, and the \gray\ and
\xray\ luminosities that may generally be expected from these events
on the basis of past experience.  Thus, a prompt high-energy emission
episode in \grbtwo\ like those seen from \sngrb\ or \xrfsn, and
occurring prior to the start of \swift\ observations, would certainly
have fallen below IPN threshold, and if present, would help to
homogenize the GRB-SNe as a class.

Finally, we note that, given that \snxrt\ manifests as an ordinary
type~Ibc SN, and that rate estimates for \snxrt-type \xray\ outbursts
are consistent with the total rate of type~Ibc supernovae
\citep{sbp+08}, it seems likely that $E_{\rm X,iso}\approx
2\tee{46}$\,erg represents a lower bound on the prompt \xray\ energy
output of any GRB-SN.


\section{Conclusions}
\label{sec:conclude}

The expanding \xray\ halos observed by \xmmlong\ in association with
\grbone\ provide a rare opportunity to study the prompt soft
\xray\ properties of a nearby GRB-supernova \citepa{vwo+04}.  We have
undertaken a detailed investigation of these halos in hopes of
clarifying the nature of the ``\xray\ blast'' \citepa{wvw+06} whose
dust-scattered \mbox{X-rays} are responsible for the halos.

In agreement with previous authors \citepa{vwo+04,whl+04,wvw+06,tm06},
we find that the properties of the halos are consistent with an
\xray\ blast occurring simultaneously with the \gray\ burst and
subsequently scattering off dust sheets at distances $d_1=871.7\pm
9.7$\,pc and $d_2=1388\pm 14$\,pc away in the plane of the Milky Way.
In addition, we discover a third expanding halo that reveals
scattering by a third dust sheet at distance $d_3 = 9.94\pm
0.39$\,kpc; the presence of this third halo is a robust feature of our
model as applied to the independent EPIC-MOS and EPIC-PN datasets, and
due to the warping of the Milky Way disk, its location is well within
the densest \HI\ regions of the outer disk.

Our constraints on the timing of the \xray\ blast, $t_0=11\pm 417$\,s
by comparison to the \grbone\ trigger time ($t_0 < 545$\,s at
90\%-confidence), are comparable in precision to the previous best
constraint \citepa{wvw+06} while suggesting a greater degree of
synchronization between the GRB and the \xray\ blast.  Combined with
upper limits from \integral\ observations of \grbone, which extend to
$t\approx 300$\,s after trigger, our findings significantly restrict
the parameter space of allowed models for the \xray\ blast.

We explore the implications of our findings, in reference to the
properties of all other spectroscopically-confirmed GRB-SNe, and
conclude that two alternative interpretations seem possible.  On the
one hand \citep{ggm+06}, the \xray\ blast may be the signature of a
high-energy ``shock breakout'' event, as observed from \grbtwo; in
this case, the event should exhibit roughly $\times$2 faster evolution
and 25\% to 200\% greater \xray\ energy than the \grbtwo\ event.
Alternatively \citepa{wvw+06}, the \xray\ blast may be the signature
of an \xray\ flare that occurred during the interval ${\rm 300\,s} <
t_0 < {\rm 545\,s}$;
if so, the ratio of \xray\ flare to \gray\ burst energies for
\grbone\ would be high but not unprecedented.

We point out that observations of \grbtwo\ by \swift\ and IPN
experiments allow for the presence of a ``precursor'' \gray\ burst as
luminous as (or more likely, somewhat less luminous than) \grbone.  In
this case, a shock-breakout origin for the \grbone\ \xray\ blast would
make \grbone\ and \grbtwo\ ``cosmic twin'' explosions with
nearly-identical high-energy properties.  If \grbone\ and
\grbtwo\ thus represent two facets of a single type of GRB-SN,
detection and observation of future nearby GRB-SNe by \swift\ can be
expected to yield additional examples of these events.


\section*{Acknowledgements}

The authors acknowledge stimulating discussions with Jamie Kennea,
Ehud Nakar, and Avishay Gal-Yam, and valuable feedback from the
anonymous referee, which has improved the paper.  
This work made use of data supplied by the UK Swift Science Data
Centre at the University of Leicester. 





\end{multicols}

\appendix

\section{Ring surface brightness}
\label{app:rings}

In this appendix we develop a semi-analytical treatment of the surface
brightness of ring-type sources as observed with \xmm.  The results
are incorporated into the numerical model of expanding dust-scattered
halos which is used in the main text to analyse the soft
\xray\ properties of \grbone.

We begin with the angularly-symmetric expression for the
\xmm\ point-spread function (PSF; Eq.~\ref{eq:psf} in the main text),
expressed as a King function:
\beq
   s_p(r) = \frac{\alpha - 1}{\pi r_c^2} \; (1 + r^2/r_c^2)^{-\alpha},
\label{eq:xmmpsf}
\eeq
where $r$ is the radial coordinate and the PSF parameters, core radius
$r_c$ and scaling index $\alpha$, are functions of the incident photon
energy and exhibit slightly different functional forms for the MOS-1,
MOS-2, and PN detectors; representative values for the parameters are
$r_c = 4.9\arcsec$ and $\alpha=1.44$.  As expressed here, the surface
brightness distribution is normalized to unit integral over
$0<r<\infty$ as long as $\alpha>1$.  We note that the \xmm\ PSF is
known not to be angularly-symmetric and that this form is adopted
strictly as a useful approximation to the actual PSF form.

\begin{figure*}
\centerline{\includegraphics[height=55mm]{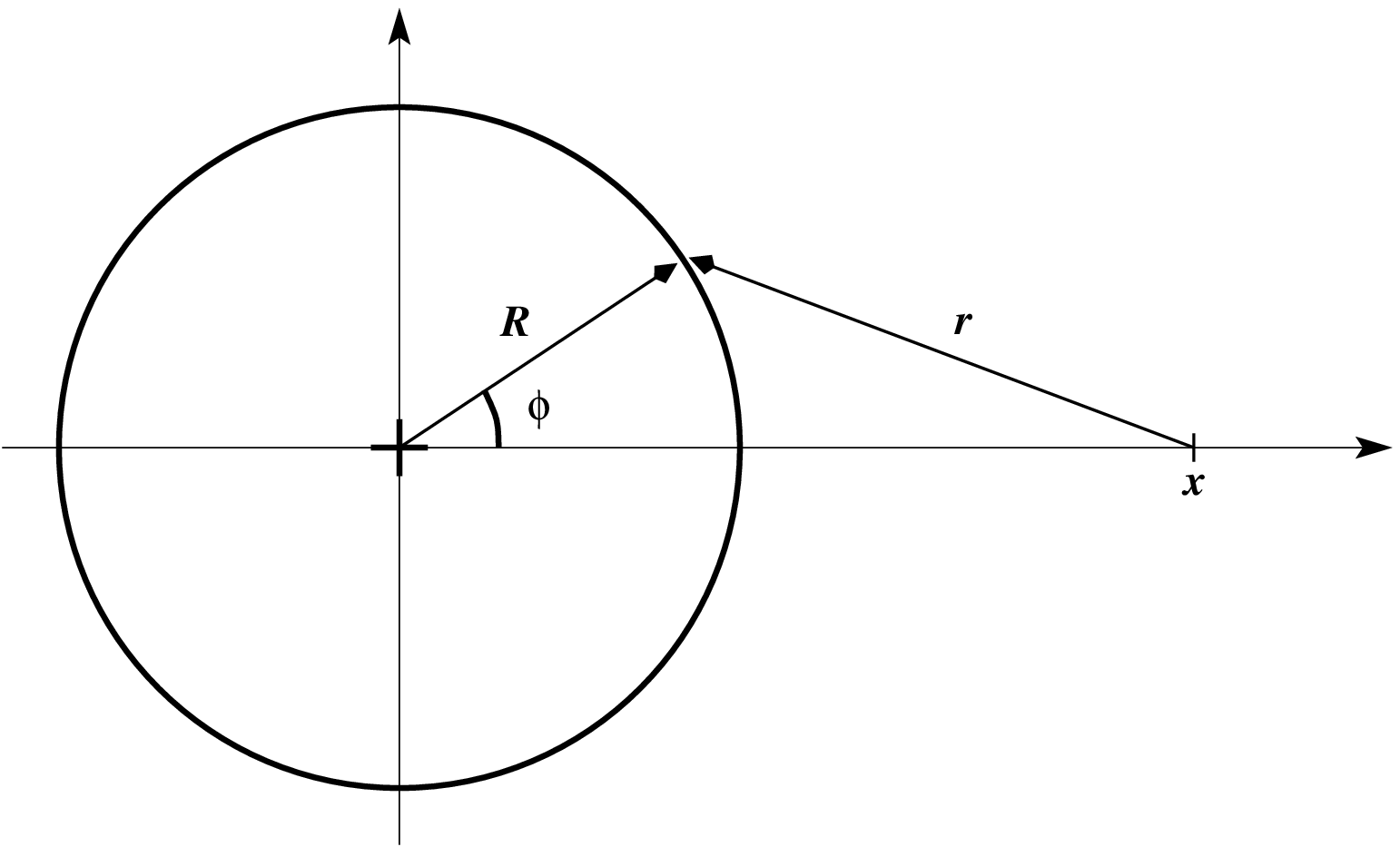}\hspace{5mm}%
            \includegraphics[height=55mm]{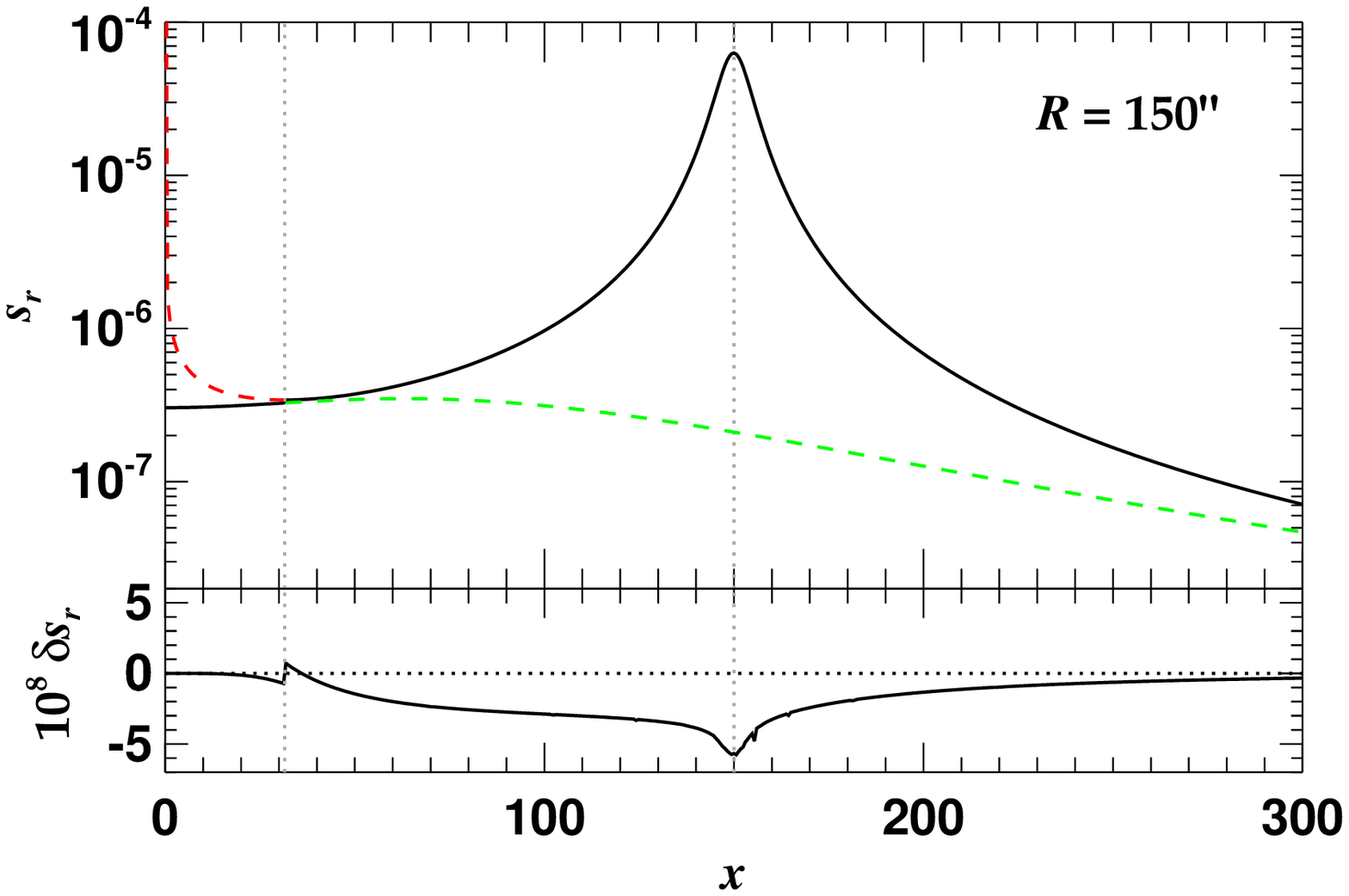}}
\caption[]{\normalsize
(left) Coordinate definitions for analytical treatment of the surface
  brightness of ring-like sources, as observed with \xmm.  The ring
  has angular radius $R$, and we seek an expression for the surface
  brightness an angular distance $x$ from the ring centre, which we
  can choose to lie on the $x$-axis, as here, without loss of
  generality in an angularly-symmetric treatment.  In analysing the
  contribution from a particular differential segment of the ring, at
  angular coordinate $\phi$, we define the distance from the segment
  position to the point $x$ to be $r$.}
\label{fig:appa}
\end{figure*}

Figure~\ref{fig:appa} serves to define the coordinate system we use in
our analysis.  We seek to describe the surface brightness at the point
$x$, due to a ring of radius $R$, as an integral over contributions
from various angles, $0\le \phi < 2\pi$, where the angular coordinate
is defined from the perspective of the ring centre.  Owing to the
angularly-symmetric nature of the PSF, we may take the point $x$ to
lie on the $x$-axis, as shown, without loss of generality.

The integral we wish to solve is thus expressed as:  
\beq
  s_r(x) = \frac{1}{2\pi} \int_0^{2\pi} 
           \frac{\alpha - 1}{\pi r_c^2} 
           (1 + r^2/r_c^2)^{-\alpha} \; d\phi,
\eeq
where the squared distance $r^2$ from the point $x$ to the ring
element at angle $\phi$ can be derived by simple trigonometry as $r^2
= R^2 + x^2 - 2Rx \cos\phi$.  Inserting this definition into the
expression for the surface brightness, and factoring out all
coefficients on the $\cos\phi$ term, we find:
\beq
  s_r(x) = \frac{\alpha - 1}{2\pi^2} \; r_c^{2(\alpha-1)} \; (2Rx)^{-\alpha}
           \int_0^{2\pi} \left(\frac{r_c^2 + R^2 + x^2}{2Rx} -
                           \cos\phi \right)^{-\alpha} \; d\phi,
\label{eq:fullint}
\eeq
where the integral is now in a form that can be solved by use of
one of the hypergeometric functions, \hyperfsym.  In particular, if we
define $A \equiv (r_c^2 + R^2 + x^2)/2Rx$, then
\beq
  \int_0^{2\pi} (A - \cos\phi)^{-\alpha} \; d\phi = 
      \pi (A-1)^{-\alpha} \;
      \hyperf{\onehalf}{\alpha}{1}{\frac{-2}{A-1}} \, + \, \\
      \pi (A+1)^{-\alpha} \; \hyperf{\onehalf}{\alpha}{1}{\frac{2}{A+1}},
\eeq
where we make use of the fact that $A>1$ in the present case (since
$r_c$, $R$, and $x$ are all positive real numbers).  The
hypergeometric function \hyperfsym\ has a series expansion that is
valid and convergent over the full domain of interest; however, for
the sake of computational speed we take two distinct approaches in our
numerical calculations.

In the first case, near the ring centre where $x\ll R$, we have $A\gg
1$, and the fourth argument of the \hyperfsym\ function in each term
of the integral solution is small.  In this case we expand each term
to second order in the fourth argument:
\beq
  \hyperf{\onehalf}{\alpha}{1}{z} = 1 + \frac{\alpha z}{2} + 
        \frac{3}{16}\, \alpha (\alpha+1) z^2 + \cdots
\eeq
so that, keeping terms to second order in $1/A$ throughout, the
overall expression for the ring surface brightness becomes: 
%
%
%
%
\beq
  s_{r2}(x) = \frac{\alpha - 1}{\pi} \; r_c^{2(\alpha-1)} \; 
             (r_c^2 + R^2 + x^2)^{-\alpha} \;
             \left[ 1 + \frac{\alpha (\alpha+1)}{4A^2} \right].  
\eeq
This is our expression for the ring surface brightness in the
second-order approximation.

For $x$ values near or beyond the ring, $x\simgt R$, with $R\gg r_c$
still holding, we take a different approach.  Writing $x=R+\Delta$, we
expand the integral expression for the ring surface brightness to
second order in $\Delta/R$:
%
%
%
%
\beq
  s_{rn}(x) = \frac{\alpha - 1}{2\pi^2} \; r_c^{2(\alpha-1)} \; (2Rx)^{-\alpha}
           \int_0^{2\pi} \left( 1 + \frac{\Delta^2}{2R^2} + 
           \frac{r_c^2}{2R^2} - \cos\phi \right)^{-\alpha} \; d\phi.
\eeq
The integrand in this expression consists of a single, strictly
positive, parenthetical term that is taken to the $(-\alpha)$ power.
In the case of the \xmm\ PSF model, $\alpha\approx 1.5 > 1$.  Thus,
the integral will be dominated by contributions from the regime where
the parenthetical term is near its minimum.  Or, to take a more
practical perspective, the bulk of the surface brightness will be
contributed by the relatively nearby portions of the ring that have
$\phi\ll 1$.  Separately, we note that the integral limits may be
changed from $0 < \phi < 2\pi$ to $-\pi < \phi < \pi$ without effect.

%
\begin{figure*}
\caption[]{\normalsize
(right) Example surface brightness profile for a ring-like source, and
  comparison of our approximate approach to a full numerical
  integration.  Upper panel: Surface brightness profile for a ring of
  radius $R=150\arcsec$ convolved with the azimuthally-symmetric
  approximation of the \xmm\ point-spread function
  (Eq.~\ref{eq:xmmpsf}).  Two distinct approximations discussed in the
  text, $s_{r2}(x)$ and $s_{rn}(x)$, are employed.  The $s_{r2}$
  approximation is used close to the ring centre ($x\le x_{\rm sep}$),
  and is plotted as a dashed green line beyond this domain.  The
  $s_{rn}$ approximation is used for $x>x_{\rm sep}$, and is plotted
  as a dashed red line near the ring centre; in the case shown,
  $x_{\rm sep}\approx 31.5\arcsec$.  Lower panel: Residuals of the
  approximate approach, as compared to a full numerical integration of
  Eq.~\ref{eq:fullint}.}
\label{fig:appb}
\end{figure*}
%

We then make two further approximations, expanding to second-order in
$\phi$, $\cos\phi\approx 1 - \onehalf\phi^2$, and increasing the
now-symmetrical limits of integration from $\pm\pi$ to $\pm\infty$.
The expression for the near-ring surface brightness then becomes:
\beq
  s_{rn}(x) = \frac{\alpha - 1}{2\pi^2} \; r_c^{2(\alpha-1)} \; (2Rx)^{-\alpha}
             \int_{-\infty}^\infty \left( 
                 \frac{\Delta^2 + r_c^2}{2R^2} + \onehalf \phi^2 
                 \right)^{-\alpha} \; d\phi,
\eeq
where we have avoided expanding $x$ to $R+\Delta$ in the $(2Rx)$ term
only.  Making a change of variables to $\psi = \phi
\sqrt{R^2/(\Delta^2 + r_c^2)}$ leads to the simplified form:  
\beq
  s_{rn}(x) = \frac{\alpha - 1}{2\pi^2} \; r_c^{2(\alpha-1)} \; 
             (2Rx)^{-\alpha} \;
             \sqrt{2} \left( \frac{\Delta^2 + r_c^2}{2R^2}
                      \right)^{-\alpha + 1/2} 
             \int_{-\infty}^\infty (1 + \psi^2)^{-\alpha} \; d\psi,
\eeq
where the integral is now in a form that can be solved via the
Euler gamma function:  
\beq
  \int_{-\infty}^\infty (1 + \psi^2)^{-\alpha} \; d\psi = 
       \frac{\sqrt{\pi}\, \Gamma(\alpha-\onehalf)}{\Gamma(\alpha)}.
\eeq
As a result, the final expression the ring surface brightness in our
``near-ring'' approximation is:  
\beq
  s_{rn}(x) = \frac{(\alpha - 1)}{\pi^{3/2} \sqrt{2}} \; r_c^{2(\alpha-1)} \; 
             (2Rx)^{-\alpha} \;
             \left( \frac{(x-R)^2 + r_c^2}{2R^2} \right)^{-\alpha + 1/2} \;
             \left( \frac{\Gamma(\alpha-\onehalf)}{\Gamma(\alpha)}
             \right).
\eeq

It remains for us to determine when to use each approximation.
In our code we use one or the other approximation depending on the
comparison of the value of $x$ to the ``separating value,'' $x_{\rm
  sep}$:
\beq
  x_{\rm sep} = \frac{2 \alpha r}{4\alpha - 1} \;
               \left( 1 - \frac{\sqrt{1 - (4\alpha - 1)(1+r_c^2/r^2)}}
                                     {2\alpha} \right),
\eeq
with the test applying as follows:
\beq
  s_r(x) = \begin{cases}
           s_{r2}(x), & x\le x_{\rm sep}; \\
           s_{rn}(x), & x > x_{\rm sep}. 
           \end{cases}
\eeq

An illustrative comparison of our numerical approach to a full
numerical integration of Eq.~\ref{eq:fullint} is shown in
Fig.~\ref{fig:appb}.  
%
%
%
%
The systematic underprediction of the true
surface brightness in the $s_{rn}$ approximation is remedied when we
renormalize the ring surface brightness over the active area of the
detector, via a numerical integration.  Overall, we find this level of
accuracy to be sufficient to our purposes; moreover, as discussed in
the main text, our use of a bootstrap Monte Carlo approach means that
any inadequacies in our modeling should result in conservative
estimates of parameter uncertainties.








\end{document}